\documentstyle[prl,aps]{revtex}

\tighten

\begin{document}

\draft   

\twocolumn

\title{Nonorthogonal quantum states maximize classical
information capacity}

\author{Christopher A.~Fuchs}

\address{Norman Bridge Laboratory of Physics 12-33, California
Institute of Technology, Pasadena, CA 91125}

\date{12 March 1997}

\maketitle

\begin{abstract}
I demonstrate that, rather unexpectedly, there exist noisy quantum
channels for which the optimal classical information transmission
rate is achieved only by signaling alphabets consisting of
nonorthogonal quantum states.
\end{abstract}


Within the framework of classical information theory, there is a
tacit but basic assumption that a communication channel's possible
inputs correspond to a set of mutually exclusive properties for the
information carriers. In the brief instant after a signal leaves the
sender's hand, but before it enters a noisy channel, an independent
observer or wire tap should be able---in principle, at least---to
read out the signal with complete reliability.  Anything less than
complete reliability in this readout represents an extra source of
noise over and above that which is supplied by the channel.
This is a situation that both the sender and receiver work to avoid.

When quantum systems are used as information carriers, one's natural
inclination is that the same basic assumption should hold. For
instance, one might think that encoding distinct signals in
nonorthogonal quantum states must be less than optimal for
information transfer. This is because the readout possibilities for
times intermediate to the signal's generation and its entrance into
the channel are excluded automatically: it is a matter of physical
law that nonorthogonal quantum states cannot be distinguished with
perfect reliability \cite{Fuchs96a} and any attempt to do so (even
imperfectly) imparts a disturbance to them \cite{Fuchs96b}.  These
are the principles that encourage the use of nonorthogonal signals
for cryptographic purposes \cite{BenBras}; however, just because of
this, one would not expect them to play a role in questions to do
with reliable, public communication.

In what follows, I present an example that dispels this
prejudice: signals encoded in {\it nonorthogonal\/} quantum states
are sometimes required to achieve the highest information transfer
rate that a channel can yield. In particular, I present a noisy
quantum mechanical channel for which the channel capacity expression
recently derived by Holevo \cite{Holevo96} and Schumacher and
Westmoreland \cite{Schumacher97} is only achieved by signals
consisting of nonorthogonal states.

In order to state the result, I first review the standard notion of
a quantum discrete memoryless channel (QDMC).
For such a channel, the information carriers are quantum systems
with a finite dimensional Hilbert space ${\cal H}_d$, $d$ denotes the
dimension.  The action of the channel is assumed to be due to
interactions between the carrier and an independent environment
outside the sender's and receiver's control. Thus, the channel's
action on the carrier's quantum state $\rho$---most generally, a
density operator---can be represented as an evolution of the form
$\rho\,\rightarrow\,\Phi(\rho)= {\rm tr}_{\scriptscriptstyle
{\rm E}}\big(U(\rho\otimes\tau)U^\dagger\big)$,
where $\tau$ denotes the standard state of the environment, $U$ is
some unitary operator, and ${\rm tr}_{\scriptscriptstyle {\rm E}}$
denotes a partial trace
over the environmental degrees of freedom.  A convenient theorem of
Kraus \cite{Kraus} is that a mapping $\Phi$ holds the form above if
and only if it can also be represented as
\begin{equation}
\rho\;\longrightarrow\;\Phi(\rho)=\sum_i A_i\rho A_i^\dagger
\label{OldMan}
\end{equation}
for some set of (possibly nonhermitian) operators $A_i$ satisfying
$\sum_i A_i^\dagger A_i=\openone$,
$\mbox{($\openone=$ the identity operator)}$.
The channel is memoryless when the evolution for arbitrary states
$\sigma$ (including entangled ones) on ${\cal H}_d^{\otimes n}$ is
\begin{eqnarray}
\Phi^{\otimes n}(\sigma)=
\sum_{i_1\cdots i_n} (A_{i_1}\otimes\cdots\otimes A_{i_n})\,\sigma\,
(A_{i_1}^\dagger\otimes\cdots\otimes A_{i_n}^\dagger)\;,
\nonumber
\end{eqnarray}
for each $n$.
That is to say, the noise acts independently on each
information carrier sent down the channel.

Let us now consider using a QDMC for the purpose of transmitting
classical information.  What we imagine here is a sender encoding
various messages $u$, $u=1,\ldots,{\cal M}$, into an equal number
of pure state preparations (i.e., one-dimensional projectors)
$\Pi_u$ on ${\cal H}_d^{\otimes n}$.
Along the way to the intended receiver, the states evolve according
to the rule above, generally emerging as mixed states
$\rho_u=\Phi^{\otimes n}(\Pi_u)$.  Finally, the receiver performs
some measurement---mathematically, a positive operator-valued measure
(POVM) \cite{Kraus}---$\{E_u\}$, with one outcome for each message
$u$. The game here is that the measurement outcome is used to
represent the receiver's best guess of the quantum state
$\rho_u$---and consequently the message $u$---appearing at the output
of the channel.  

Note that the formulation so far is completely general in its usage
of the QDMC.  In particular, the quantum states used to encode the
messages may be massively entangled across the $n$ transmissions
\cite{BFS}.  Moreover, the POVM $\{E_u\}$ may be
a collective quantum measurement over the whole Hilbert space
${\cal H}_d^{\otimes n}$, and need not factorize into measurements
on the individual carriers \cite{Holevo79}.  For the considerations
here, however, I restrict attention to senders using encodings based
on a {\it finite alphabet}.  A sender is said to make use of a finite
alphabet when his signals are restricted to be product states on
${\cal H}_d^{\otimes n}$, all of which are drawn from some fixed
finite set ${\cal X}=\{\Pi_\ell\}$, $\ell=1,\ldots,m$, of
pure states on ${\cal H}_d$.  That is to say, the
sender is now imagined to encode messages $u=(\ell_1,\ldots,\ell_n)$
into quantum states of the form
$\Pi_u=\Pi_{\ell_1}\otimes\cdots\otimes\Pi_{\ell_n}$.  Such an
encoding, taken as a whole, is called a {\it code}.

With this, we can turn to the issue of reliable transmission of
information through the channel. A
$(\lfloor2^{nR}\rfloor,n,\lambda_n)$ code, $0\le R\le1$, is a set of
$\lfloor2^{nR}\rfloor$ codewords $\Pi_u$ (each of length $n$) such
that the maximum probability of error in guessing a message is
$\lambda_n$, i.e.,
$\lambda_n=\max_u\big(1-{\rm tr}(\rho_u E_u)\big)$.
The number $R$ appearing in this definition is known as the
$\it rate\/$ of information transfer of the code. A rate $R$ is said
to be achievable if there exists a sequence of
$(\lfloor2^{nR}\rfloor,n,\lambda_n)$ codes with
$\lambda_n\rightarrow 0$ as
$n\rightarrow\infty$. The {\it capacity\/} $C$ of the QDMC is the
supremum of all achievable rates, where the supremum is taken
explicitly over all alphabets used for coding, all codes making
use of that alphabet, and all possible POVMs used for decoding at
the receiver.  Our main concern here is in finding
the optimal alphabet for the encoding, the issue being whether the
optimal alphabet must consist of orthogonal states or not.

A method of calculating the capacity has been known for some time
when the POVM elements $E_u$ are, like the codewords in this
scenario, restricted to
be tensor product operators on ${\cal H}_d^{\otimes n}$
\cite{Holevo73}. This restriction is equivalent to saying that
collective measurements on codewords are excluded from the
game; each information carrier is measured individually. The
restricted capacity $C_1$ is given by the supremum {\it accessible
information\/} $I_1({\cal E})$
\cite{Fuchs96a} over all signal ensembles ${\cal E}=\{p_i;\Pi_i\}$,
$p_i\ge0$, $\sum_i p_i=1$; i.e.,
\begin{equation}
C_1=\sup_{\cal E} I_1({\cal E})\;,
\label{Lambic}
\end{equation}
where
\begin{equation}
I_1({\cal E})= \sup_{\{E_b\}}
\Big[H\big({\rm tr}(\rho E_b)\big)-
\sum_i p_i H\big({\rm tr}(\rho_i E_b)\big)\Big]\;,
\label{Icarus}
\end{equation}
$\rho_i=\Phi(\Pi_i)$ are the output states, $\rho=\sum_i p_i\rho_i$,
and $H\big({\rm tr}(\tau E_b)\big)=-\sum_b {\rm tr}(\tau E_b)
\log {\rm tr}(\tau E_b)$ is the Shannon entropy for the
probability distribution ${\rm tr}(\sigma E_b)$ derived from a POVM
$\{E_b\}$.  (All logarithms throughout are calculated base 2.)

Expression~(\ref{Lambic}) coincides with the standard classical
capacity theorem of Shannon \cite{Shannon48} for a discrete
memoryless channel: it is just that in the quantum case extra care
must be taken to optimize both the input alphabet and the output
observable---neither is given {\it a priori}.  Note that the
supremization in Eq.~(\ref{Icarus}) is over {\it all\/}
POVMs on ${\cal H}_d$: for this expression there is no restriction
that the number of POVM elements be the same as the number of states
in the alphabet $\cal X$. However, convexity arguments can be used
to show that Eqs.~(\ref{Lambic}) and (\ref{Icarus}) are achievable
by ensembles and POVMs {\it each\/}
with no more than $d^2$ elements \cite{Davies78,Fujiwara97}.

Recently, an elegant expression for the capacity $C$ has
been derived \cite{Holevo96,Schumacher97}, which dispenses with
an explicit optimization over the receiver's
measurement.  The theorem is that
\begin{equation}
C = \sup_{\cal E} I({\cal E})
\label{ShimmyShake}
\end{equation}
where
\begin{equation}
I({\cal E}) = S(\rho)-\sum_i p_i S(\rho_i)\;,
\label{BellyRoll}
\end{equation}
$\rho_i$ and $\rho$ are defined as above, and
$S(\tau)=-{\rm tr}(\tau\log\tau)$ is the von Neumann entropy
of a density operator $\tau$.  Here again, convexity arguments
\cite{Fujiwara97,Uhlmann97} give that the supremum can be achieved by
signal ensembles consisting of no more than $d^2$ terms.

It is important to note that, depending upon the channel, $C$ can
be strictly greater than $C_1$.  This is a result of the fact that
collective measurements generally afford more power for
distinguishing product states than do product measurements
\cite{Holevo79,Peres91,Hausladen96}.  Moreover, this point is
doubly significant for the task at hand because collective
measurements also appear to be the key for eliciting the optimality
of nonorthogonal inputs.

With the Theorems (\ref{Lambic}) and (\ref{ShimmyShake}) for the
capacities $C_1$ and $C$ in hand, the
last remark can be made precise.  The question is this.  Do there
exist channels for which Eq.~(\ref{ShimmyShake}) is achieved
{\it only\/} by an ensemble of nonorthogonal states?  I will answer
this in the affirmative by explicitly constructing an example of a
channel on ${\cal H}_2$ that requires, at the very least, a
nonorthogonal binary alphabet to achieve capacity.  That is to say,
I shall exhibit a particular $\Phi$ and a particular ensemble
${\cal E}_p=\{\frac{1}{2},\frac{1}{2};\Pi_0,\Pi_1\}$ with
${\rm tr}(\Pi_0\Pi_1)\ne0$ for which
$C\ge I({\cal E}_p)> \sup_{{\cal E}_\perp} I({\cal E}_\perp)$.
The rightmost supremization in this is taken exclusively
over ensembles of orthogonal states.

As stated earlier, this situation is somewhat surprising.
Indeed it can be shown for general $\Phi$ and ${\cal H}_d$
that when the issue is of distinguishing
two outputs in an optimal way---rather than optimizing information
rate---{\it and\/} there are no restrictions on the inputs or the
POVMs, then orthogonal inputs are always sufficient \cite{BFS}.
Moreover, when $d=2$ and the input alphabet is binary, the capacity
$C_1$ is always achievable by an orthogonal alphabet: this will be
demonstrated later in the paper.  For the present, I turn to the
particular example.

The ``splaying'' channel acting on density operators of ${\cal H}_2$
is described simply enough by means of a Kraus representation as in
Eq.~(\ref{OldMan}).  The $A_i$ used to define it are
\begin{eqnarray}
A_x=\sqrt{\frac{2}{3}}|x\rangle\langle x|\,,
\quad
A_y=\sqrt{\frac{2}{3}}|y\rangle\langle +|\,,
\quad
A_{\overline{y}}=\sqrt{\frac{2}{3}}|\overline{y}\rangle\langle -|\,,
\nonumber
\end{eqnarray}
where, fixing an orthonormal basis
$\{|x\rangle,|\overline{x}\rangle\}$ on ${\cal H}_2$,
\begin{eqnarray}
|y\rangle=\frac{1}{\sqrt{2}}\big(|x\rangle+|\overline{x}
\rangle\big)\;,
&\quad\quad&
|\overline{y}\rangle=\frac{1}{\sqrt{2}}\big(|x\rangle-
|\overline{x}\rangle\big)\;,
\\
|+\rangle=\frac{1}{2}|x\rangle+\frac{\sqrt{3}}{2}
|\overline{x}\rangle\;,
&\quad\quad&
|-\rangle=\frac{1}{2}|x\rangle-\frac{\sqrt{3}}{2}
|\overline{x}\rangle\;.
\end{eqnarray}

The action of this channel can be thought of in more graphic terms
as follows.  Let us make a switch to Bloch-sphere notation for all
operators.  The channel, personified as Eve, begins by performing
the symmetric three-outcome ``trine'' POVM
as the quantum states make their way from sender to
receiver. I.e., the positive operators in her POVM are given by
\begin{equation}
E_i = \frac{1}{3}(\openone + \bbox{n}_i\cdot\bbox{\sigma}),
\end{equation}
where $\bbox{\sigma}$ is the vector of Pauli matrices,
$\bbox{n}_x=(1,0,0)$, and $\bbox{n}_\pm = (-1/2,\pm\sqrt{3}/2,0)$.
The three vectors here are 120 degrees apart and confined to the
$x$-$y$ plane; as must be the case for all POVMs,
$E_x+E_++E_-=I$. Upon receiving outcome $i$, Eve forwards a quantum
state $\eta_i$ to Bob according to the rule
\begin{equation}
\eta_x = \frac{1}{2}(\openone+\bbox{x}\cdot\bbox{\sigma})
\quad\mbox{and}\quad
\eta_\pm = \frac{1}{2}(\openone\pm\bbox{y}\cdot\bbox{\sigma})\;,
\end{equation}
where $\bbox{x}=(1,0,0)$ and $\bbox{y}=(0,1,0)$.  The key idea is that
if $E_x$ is detected, the state corresponding to the outcome is
forwarded to the receiver; however, if $E_+$ or $E_-$ are detected,
orthogonal or ``splayed'' versions of the outcomes are sent.

If the sender transmits a general pure state
\begin{equation}
\Pi_{\alpha\beta} = \frac{1}{2}(\openone+\bbox{s}_{\alpha\beta}\cdot
\bbox{\sigma})\;,
\label{DonaldDuck}
\end{equation}
where $\bbox{s}_{\alpha\beta}=
(\cos\alpha\sin\beta\,,\,\sin\alpha\sin\beta\,,\,\cos\beta),$
for $\alpha\in[0,2\pi)$ and $\beta\in[0,\pi)$,
the upshot of Eve's interference---as far as the sender and
receiver are concerned---is the evolution
$\Pi_{\alpha\beta}\,\rightarrow\,\Phi(\Pi_{\alpha\beta})$
where
\begin{equation}
\Phi(\Pi_{\alpha\beta})
\,=\,
\sum_i {\rm tr}(\Pi_{\alpha\beta} E_i)\,\eta_i
\,=\,
\frac{1}{2}(\openone+\bbox{t}_{\alpha\beta}\cdot\bbox{\sigma})
\label{paunch}
\end{equation}
and
$\bbox{t}_{\alpha\beta}=
\frac{1}{3}\big(1+\cos\alpha\sin\beta\,,\,
\sqrt{3}\sin\alpha\sin\beta\,,\,0\big)$. This follows since
${\rm tr}(\Pi_{\alpha\beta} E_x)=(1+\cos\alpha\sin\beta)/3$ and also
${\rm tr}(\Pi_{\alpha\beta} E_\pm)=\big(2-
\cos\alpha\sin\beta \pm \sqrt{3}\sin\alpha\sin\beta\big)/6$.

With Eq.~(\ref{paunch}), one can readily calculate
Eq.~(\ref{BellyRoll}) for an arbitrary ensemble of {\it orthogonal\/}
input states. Suppose the state in Eq.~(\ref{DonaldDuck}) and one
orthogonal to it (i.e., with Bloch vector
$-\bbox{s}_{\alpha\beta}$) are sent through the channel with prior
probabilities $t$ and $1-t$, respectively. Calling the result
of Eq.~(\ref{BellyRoll}) $I(\alpha,\beta,t)$, this gives
\begin{eqnarray}
&&
I(\alpha,\beta,t)=
\nonumber\\
&&
\phi\!\left[
\big(1+(2t-1)\cos\alpha\sin\beta\big)^2 +
3\big((2t-1)\sin\alpha\sin\beta\big)^2\,\right]
\nonumber\\
&&
-t\,\phi\!\left[
\big(1+\cos\alpha\sin\beta\big)^2 +
3\big(\sin\alpha\sin\beta\big)^2\,\right]
\nonumber\\
&&
-(1-t)\,
\phi\!\left[
\big(1-\cos\alpha\sin\beta\big)^2 +
3\big(\sin\alpha\sin\beta\big)^2\,\right]
\label{XFiles}
\end{eqnarray}
where $\phi(x)=-h(\sqrt{x}/3)$ and
\begin{equation}
2\,h(z)=(1+z)\log(1+z)+(1-z)\log(1-z)\;.
\end{equation}
One can easily check that Eq.~(\ref{XFiles}) is maximized when
$\alpha=\beta=\pi/2$ and $t=1/2$, yielding a value of
\begin{equation}
C_{\mbox{ortho}}=
\frac{1}{6}\log\!\left(\frac{3125}{1024}\right)
\approx 0.268273 \mbox{ bits}.
\end{equation}

Now consider the following ensemble of inputs.  Let
$\Pi_\alpha$ be a state given by Eq.~(\ref{DonaldDuck})
but with $\beta=\pi/2$, and let 
$\overline{\Pi}_\alpha=\Pi_{-\alpha}$. Assume
each of these occurs with prior probability $1/2$.  Thus,
the two signaling states in this ensemble are (generally)
nonorthogonal, but restricted to the plane of the POVM elements
and reflecting their symmetry.  Again, one readily calculates
Eq.~(\ref{BellyRoll}) to get
\begin{eqnarray}
I(\alpha)=\phi\big((1+\cos\alpha)^2\big) -
\phi\big((1+\cos\alpha)^2 + 3\sin^2\alpha\big)\;.
\label{Bertis}
\nonumber
\end{eqnarray}
The analytic maximization of this quantity depends upon
the solution of a transcendental equation.  Therefore, the
maximization requires some numerical work: it
turns out to be attained when $\alpha=1.521808\ne\pi/2$,
roughly 87.2 degrees. The value of the maximum is
\begin{equation}
C_{\mbox{nono}}\approx0.268932\mbox{ bits}.
\label{Eureka}
\end{equation}
This completes the demonstration that a QDMC's classical
information capacity need not be achievable by orthogonal states.
The difference in this particular example is not large, but it is
enough to prove the principle.

Heuristically, what is going on with the splaying channel is that,
from a ``God's eye view,'' the output $\eta_x$ acts like an erasure
flag, signifying the disappearance of a bit. As the angle $\alpha$ is
reduced, the probability of a flagged erasure increases, and so the
information rate decreases.  As $\alpha$ is made larger, the
transmission probability for distinguishable bits (i.e., $\eta_+$ and
$\eta_-$) increases, but there is an accompanying increased
probability that a bit will have flipped.  The angle $\alpha$ in
Eq.~(\ref{Eureka}) represents the optimal tradeoff between these
tensions, as quantified by the capacity formula for $C$ in
Eq.~(\ref{ShimmyShake})---in other words, when the full power of
collective quantum measurements is made available at the receiver.

The last point appears to be crucial for understanding the origin of
this effect.  When each qubit is measured individually, the optimal
tradeoff between the tensions is quantified by the capacity $C_1$
given in Eq.~(\ref{Lambic}).  In that case, it should be noted that
the erasure flag's contribution to the tensions effectively
disappears; with respect to individual measurements, the erasure
flag always manifests itself as a probability for a bit flip error.
This is seen easily with an example.
If the ensemble $\{\Pi_\alpha,\overline{\Pi}_\alpha\}$ (equal prior
probabilities) is used, but no collective measurements, then it turns
out that there is enough symmetry in the problem that
Eq.~(\ref{Icarus}) can be calculated explicitly.  When two
equiprobable states with equal-length Bloch vectors
$\bbox{a}$ and $\bbox{b}$ are to be distinguished, the optimal
measurement is specified by the unit vectors parallel and
anti-parallel to $\bbox{d}=\bbox{a}-\bbox{b}$, and the accessible
information is given by 
$I_1(\bbox{a},\bbox{b})=-\phi\big(9(\bbox{a}\cdot\bbox{d})/2\big)$
\cite{Fuchs96a,Levitin}.
For the case at hand, we obtain $I_1(\alpha)=\phi(3\sin^2\alpha)$,
which is achieved by a measurement basis consisting of the projectors
$\eta_+$ and $\eta_-$.  As far as this measurement is concerned,
an erasure-flag output has equal probabilities for leading to correct
and incorrect identifications by the receiver.  In particular,
$I_1(\alpha)$ has a maximum of 0.255992 bits at $\alpha=\pi/2$, i.e.,
for an ensemble of orthogonal input states.  

Indeed, it is a generic property of channels on ${\cal H}_2$ with
{\it binary\/} input alphabets that the maximum achievable rate with
respect to individual measurements can be attained by an orthogonal
alphabet.  Furthermore, since a standard orthogonal
projection-valued measurement always suffices
for achieving capacity here \cite{Fuchs96b,Levitin}, this remains
true even without optimizing the ensemble prior probabilities or the
measurement observable. This fact arises in the following manner.

Suppose a fixed measurement is given by the Bloch vectors
$\bbox{n}$ and $-\bbox{n}$, and the binary signal alphabet (yet to
be optimized) is associated with fixed prior probabilities $1-t$ and
$t$ ($t\ne0,1$) for its letters.  Let $\bbox{a}$ and $\bbox{b}$
denote the respective Bloch vectors associated with the signal
alphabet, and let $\bbox{c}=(1-t)\bbox{a}+t\bbox{b}$.  The effect of
the channel on these Bloch vectors is to transform them according to
some affine transformation \cite{Fujiwara97}:
$\bbox{a}\rightarrow \bbox{a}^\prime=M\bbox{a}+\bbox{e}$,
$\bbox{b}\rightarrow \bbox{b}^\prime=M\bbox{b}+\bbox{e}$,
etc., where $M$ is a real $3\times3$ matrix and $\bbox{e}$ is a
fixed vector within the Bloch sphere.  With these notations, the
mutual information $J$ between input and output for this ensemble is
\begin{eqnarray}
J
&=&
-h\big(\bbox{c}^\prime\cdot\bbox{n}\big)+
(1-t)h\big(\bbox{a}^\prime\cdot\bbox{n}\big)+
t\,h\big(\bbox{b}^\prime\cdot\bbox{n}\big)
\nonumber\\
&=&
-h(\bbox{c}\cdot\tilde{\bbox{n}}+w)+
(1-t)h(\bbox{a}\cdot\tilde{\bbox{n}}+w)+
th(\bbox{b}\cdot\tilde{\bbox{n}}+w),
\nonumber
\end{eqnarray}
where $\tilde{\bbox{n}}=M^{\rm T}\bbox{n}$ and
$w=\bbox{e}\cdot\bbox{n}$.  A necessary condition on any candidates
$\bbox{a}$ and $\bbox{b}$ for optimizing this mutual information is
that it be invariant to first order with respect to small variations
about these vectors.  Taking into account the constraint that the
inputs be pure states, this leads to the following two variational
equations:
\begin{eqnarray}
\bbox{0}
&=&
\log\!\left[\frac{\big(1-w-\bbox{c}\cdot\tilde{\bbox{n}}\big)
\big(1+w+\bbox{a}\cdot\tilde{\bbox{n}}\big)}
{\big(1+w+\bbox{c}\cdot\tilde{\bbox{n}}\big)
\big(1-w-\bbox{a}\cdot\tilde{\bbox{n}}\big)}\right]\bbox{n}_a
\\
\bbox{0}
&=&
\log\!\left[\frac{\big(1-w-\bbox{c}\cdot\tilde{\bbox{n}}\big)
\big(1+w+\bbox{b}\cdot\tilde{\bbox{n}}\big)}
{\big(1+w+\bbox{c}\cdot\tilde{\bbox{n}}\big)
\big(1-w-\bbox{b}\cdot\tilde{\bbox{n}}\big)}\right]\bbox{n}_b\;,
\end{eqnarray}
where $\bbox{0}$ is the zero vector,
$\bbox{n}_a=\tilde{\bbox{n}}-\big(\tilde{\bbox{n}}\cdot\bbox{a}\big)
\bbox{a}$, and
$\bbox{n}_b=\tilde{\bbox{n}}-\big(\tilde{\bbox{n}}\cdot\bbox{b}\big)
\bbox{b}$.  It is easy to check that the only solutions to
these occur when either $\bbox{a}=\bbox{b}$,
$\tilde{\bbox{n}}\cdot\bbox{a}=\tilde{\bbox{n}}\cdot\bbox{b}$, or
$\bbox{a}=-\bbox{b}$.  In the first two cases the mutual information
vanishes; in the last case, it is maximal.  This proves the point.

In summary, I have shown that, contrary to some intuition,
there exist noisy quantum channels for which nonorthogonal input
states lead to the largest reliable information transfer rate. In
the particular example here, and indeed for all possible
channels on ${\cal H}_2$, collective measurements appear to play a
crucial role in bringing about the effect {\it whenever it exists}.
However, it remains an open question whether {\it collective
measurements following product-state inputs} is the one and only
ingredient required for bringing about the optimality of
nonorthogonal inputs:  for instance, it is not known whether there
exists a channel on ${\cal H}_d$, $d\ge3$, for which the capacity
$C_1$ is only attained for a nonorthogonal input alphabet.

The particular example exhibited here was somewhat contrived,
being built explicitly to show the desired effect.  However, since the
completion of this work, several ``real world'' channels have
been discovered (through numerical simulation) to require
nonorthogonal inputs to achieve capacity.  In fact, the effect appears
to be generic for channels of a certain dissipative character---the
standard amplitude damping channel being one such example.  An
extended discussion of these channels will appear elsewhere
\cite{Fuchs97}.

I thank H.~Barnum, C.~Bennett, H.~Mabuchi, P.~Shor,
J.~Smolin, and A.~Uhlmann for helpful discussions. This work was
supported by a Lee A. DuBridge Fellowship and by DARPA through the
Quantum Information and Computing (QUIC) Institute administered by
ARO.


\end{document}